\documentclass{article}

\usepackage[a4paper,margin=0.8in]{geometry}
\usepackage{graphicx}
\usepackage{epsfig}
\usepackage{amsmath,amssymb}
\usepackage{bm}
\usepackage{xcolor}
\usepackage{pdflscape}
\usepackage{color}
\usepackage{cancel}
\usepackage{units}

\usepackage{psfrag}
\usepackage{mathrsfs}
\usepackage{esdiff}

\newcommand{\erre}{{\bf r}}
\newcommand{\erredot}{\dot{\bf r}}
\newcommand{\angmom}{{\bf c}}
\newcommand{\energy}{\mathcal{E}}
\newcommand{\kepler}{\mathcal{K}}
\newcommand{\lenz}{{\bf L}}
\newcommand{\lambert}{\mathcal{L}}
\newcommand{\bX}{{\bf X}}

\newcommand{\bDelta}{\bm{\Delta}}
\newcommand{\bE}{\mathbf{E}}

\def\alphadot{\dot{\alpha}}
\def\deltadot{\dot{\delta}}
\def\da{\Delta\alpha}
\def\dd{\Delta\delta}

\def\erho{{\bf e}^\rho}
\def\ealpha{{\bf e}^\alpha}
\def\edelta{{\bf e}^\delta}

\def\bq{{\bf q}}
\def\br{{\bf r}}

\def\rhodot{\dot{\rho}}
\def\qalpha{\bq\cdot\ealpha}
\def\qdelta{\bq\cdot\edelta}                    
\def\dq{\dot{\bf q}} 
\def\dqalpha{\dq\cdot\ealpha}
\def\dqdelta{\dq\cdot\edelta}

\def\bv{{\bf v}}
\def\by{{\bf y}}
\def\bzero{{\bf 0}}

\def\R{\mathbb{R}}

\def\N{\mathbb{N}}

\def\elatt{\mathscr{E}_{att}}

\newtheorem{remark}{\bf Remark}

\title{Preliminary orbits with line-of-sight
  correction for LEO satellites observed with radar}

\author{H. Ma$^{1}$\thanks{E-mail: helenema@gmail.com},
  G. F. Gronchi$^{1}$,
  D. Bracali Cioci$^{2}$
  \\ $^{1}$Dipartimento di Matematica, Universit\`a di Pisa, Largo
  B. Pontecorvo 5, Pisa 56127, Italy\\ $^{2}$Space Dynamics Services
  s.r.l., Via M. Giuntini 63, Navacchio 56023, Italy}

\begin{document}

\maketitle

\begin{abstract}
We propose a method to account for the Earth oblateness effect in
preliminary orbit determination of satellites in low orbits with radar
observations.
This method is an improvement of the one described in \cite{gdbm15},
which uses a pure Keplerian dynamical model.  Since the effect of the
Earth oblateness is strong at low altitudes, its inclusion in the
model can sensibly improve the initial orbit, giving a better starting
guess for differential corrections and increasing the chances to
obtain their convergence.
The input set consists of two tracks of radar observations, each one
composed of at least 4 observations taken during the same pass of the
satellite.
A single observation gives the topocentric position of the satellite,
where the range is very accurate, while the line of sight direction is
poorly determined.
From these data we can compute by a polynomial fit the values of the
range and range rate at the mean epochs of the two tracks.
In order to obtain a preliminary orbit we wish to compute the angular
velocities, that is the rate of change of the line of sight. In the
same spirit of \cite{gdbm15}, we also wish to correct the values of
the angular measurements, so that they fit the selected
dynamical model if the same holds for the radial distance and velocity.
The selected model is a perturbed Keplerian dynamics, where the only
perturbation included is the secular effect of the $J_2$ term of
the geopotential.  The proposed algorithm models this problem with 8
equations in 8 unknowns.
\end{abstract}

\medskip


\section{Introduction}
We investigate a method of preliminary orbit determination that takes
into account the Earth oblateness effect. The method is conceived for
the computation of orbits of Earth satellites at low altitudes
(LEO), with radar observations. This is an improvement of the
algorithm described in \cite{gdbm15}, where a two-body approximation
is employed.
The Earth oblateness has been already included in the dynamical model for
the computation of preliminary orbits, see e.g. \cite{ftmr10},
where radar observations were assumed to measure the
range, the range rate, the right ascension and the declination.
In this paper we include the secular effect of the $J_2$ term of the
geopotential in the dynamical model, and we aim to compute preliminary
orbits where also corrections to the values of the angular
measurements are applied.  To write the equations for this problem we
shall use the Keplerian integrals evolution, the equations of motion
projected onto the line of sight, and Lambert's theorem.
%
We assume that each radar measurement at epoch $t$ is composed by a
precise value of the range $\rho$ (with standard deviation $\approx 10$
m) and poorly determined values of right ascension $\alpha$ and
declination $\delta$ (with standard deviation $\approx 0.2$ deg), as
we did in \cite{gdbm15}.
The available data are radar tracks of the form
\begin{equation}
\hskip 2cm 
(t_j,\rho_j,\alpha_j,\delta_j), \qquad j=1,\dots , 4
\label{radtrail}
\end{equation}
where $\Delta t = t_{j+1}-t_j \approx 10$ s.
%
Given a radar track we can derive the vector
\begin{equation}
(\bar t, \bar\alpha, \bar\delta, \rho, \dot\rho, \ddot\rho),
\label{radinterp}
\end{equation}
where $\bar t$, $\bar\alpha$ and $\bar\delta$ are the mean values and
$\rho$, $\dot\rho$, $\ddot\rho$ are obtained through a
cubic fit.

To describe the osculating two-body orbit of the satellite we use
spherical coordinates (known as {\em attributable} coordinates)
\begin{equation}
\mathscr{E}_{att}=(\alpha,\delta,\alphadot,\deltadot,\rho,\dot\rho).
\label{attel}
\end{equation}
Given the data in (\ref{radinterp}), we miss the values of
$\alphadot,\deltadot$ to have an orbit, which are the unknowns of our
orbit determination problem.
Moreover, to improve the quality of the preliminary orbit we wish to
correct the values of $\bar\alpha, \bar\delta$.  For this purpose
we introduce the quantities $\da, \dd$,  that are unknown small
deviations from the mean values $\bar{\alpha}$, $\bar{\delta}$:
\[
\alpha = \bar\alpha + \da,\hskip 1cm \delta = \bar\delta + \dd.
\]
We call {\em infinitesimal angles} the deviations $\da, \dd$ . 
Moreover, in place of the unknowns $\alphadot, \deltadot$ we use
\[
\xi = \rho\alphadot\cos\delta, \hskip 1cm \zeta = \rho\deltadot,
\]
that are the components of the topocentric velocity of the satellite
orthogonal to the line of sight.  The orbit at time $\bar{t} -
\rho/c$, where $c$ is the velocity of light,
is completely determined by the coordinates
\[
\mathscr{E}_{att}^*=(\da,\dd,\xi,\zeta,\rho,\dot\rho),
\]
that we call {\em modified attributable} coordinates.
Our method consists in correlating two radar tracks to determine the
values of the unknowns and compute preliminary orbits, see \cite{mg10}. 
Therefore,  the input data set is
\begin{equation}
(\bar t_1, \bar\alpha_1, \bar\delta_1, \rho_1, \dot\rho_1,
  \ddot\rho_1), \hskip 1cm (\bar t_2, \bar\alpha_2, \bar\delta_2,
  \rho_2, \dot\rho_2, \ddot\rho_2).
\label{radinput}
\end{equation} 
We want to determine the values of the vectors
$(\da_1,\dd_1,\xi_1,\zeta_1)$ and $(\da_2,\dd_2,\xi_2,\zeta_2)$ at
epochs $\bar{t}_1-\rho_1/c$ and $\bar{t}_2-\rho_2/c$.
%


\section{Notation}
\label{s:notation}

Let us denote by $\erho$ the unit vector corresponding to the line of
sight, and by $\bq$ the geocentric position of the observer. Then the
geocentric position of the observed body is
\begin{equation}
  \br = \bq + \rho\erho,
  \label{geocpos}
\end{equation}
where $\rho$ is the range. Using as angular coordinates the
topocentric right ascension $\alpha$ and declination $\delta$ in an
equatorial reference frame (e.g. J2000), we have
\[
\erho = (\cos\delta\cos\alpha, \cos\delta\sin\alpha, \sin\delta).
\]
We introduce the unit vectors
\[
\ealpha = 
\left(- \cos\delta \sin\alpha, \cos\delta \cos\alpha,0 \right)
,
\quad
\edelta = 
\left(-\sin\delta \cos\alpha,-\sin\delta \sin\alpha,\cos\delta \right).
\]
The set $\{\erho,\ealpha,\edelta\}$ is an orthonormal system.
Denoting by $\dot\erre$  
the geocentric velocity of the satellite, we have
\begin{equation}
  \dot\br = \xi\ealpha + \zeta\edelta + (\dot\rho\erho + \dot\bq).
  \label{geocvel}
\end{equation}

We will use the following different sets of coordinates for the orbits: 
\begin{align*}
  \mathscr{E}_{kep}  &= (a,e,I,\Omega,\omega,\ell), \\
  \mathscr{E}_{car}  &= (x,y,z,\dot{x},\dot{y},\dot{z}), \\
  \mathscr{E}_{att}  &= (\alpha,\delta,\alphadot,\deltadot,\rho,\rhodot),\\
  \mathscr{E}^*_{att}&= (\Delta \alpha,\Delta \delta,\xi,\zeta,\rho,\rhodot),
\end{align*}
that are respectively Keplerian, Cartesian, attributable and modified
attributable coordinates.

We also consider the coordinate changes
\begin{equation*}
  \displaystyle \mathscr{E}_{kep} \xrightarrow[]{ \displaystyle\phi_1}
 \mathscr{E}_{car},
\hskip 1cm 
 \mathscr{E}_{car}  \xrightarrow[]{ \displaystyle\phi_2}
  \mathscr{E}_{att},
\hskip 1cm
 \mathscr{E}_{att}  \xrightarrow[]{ \displaystyle\phi_3}
  \mathscr{E}^*_{att},
\end{equation*} 
and the composite transformation
\[
\Phi = \phi_3 \circ \phi_2 \circ \phi_1
\] 
from $\mathscr{E}_{kep}$ to $\mathscr{E}^*_{att}$.

\section{The equations of motion}
\label{s:eqmotion}

Let us consider Newton's equation
\begin{equation}
\ddot{\erre} = \nabla U(\erre),
\label{J2eq}
\end{equation}
for the motion of a point mass in the Earth gravity field where the
potential $U$ is truncated at the $J_2$-term, that is
\begin{equation}
U (\erre) = \frac{\mu}{r}\Bigl[1 - J_2\Bigl(\frac{R_\oplus}{r}\Bigr)^2
  P_2(\sin\delta)\Bigr] .
\end{equation}
Here, $r=|\erre|$ is the geocentric distance, $R_\oplus$ is the
equatorial radius of the Earth and $P_2$ is the Legendre polynomial of
second degree
\[
P_2(\sin\delta)=\frac{3}{2}\sin^2\delta - \frac{1}{2} =
\frac{3}{2}\frac{z^2}{r^2} - \frac{1}{2}.
\]
The problem defined by equation \eqref{J2eq} is non-integrable, see
\cite{CN}.  If we average out the short period term in \eqref{J2eq} we
obtain an integrable system (see \cite{Roy}) given by
\begin{equation}
  \begin{cases}
    \quad \dot{a} = 0 \\
    \quad \dot{e} = 0 \\
    \quad \dot{I} = 0 \\
    \quad \dot{\Omega} = 
\displaystyle -\frac{3}{2}J_2\frac{R_{\oplus}^2}{p^2}n_0\cos{I} \\
    \quad \stackrel{}{\dot{\omega}} = 
\displaystyle \frac{3}{4}J_2\frac{R_{\oplus}^2}{p^2}n_0(4-5 \sin^2{I}) \\
    \quad  \stackrel{}{\dot{\ell}}   = n = 
\displaystyle n_0 \Bigl[1+\frac{3}{2}J_2\frac{R^2_{\oplus}}{p^2}
      \Bigl(1-\frac{3}{2}\sin^2 I\Bigr) \sqrt{1-e^2}\Bigr]
  \end{cases}
\label{J2eqav}
\end{equation}
with $p=a(1-e^2)$ the parameter of the two-body trajectory and
$n_0=\sqrt{\mu/a^3}$ the mean motion.
Note that, in the dynamics defined by (\ref{J2eqav}), the elements
$a,e,I$ remain constant while the ascending node $\Omega$, the
argument of perigee $\omega$, and the mean anomaly $\ell$ change
uniformly with time.
Equations (\ref{J2eqav}) can be written shortly as
\begin{equation}
\dot{\mathscr{E}}_{kep} = \bX_{kep}(\mathscr{E}_{kep}).
\label{J2eqavshort}
\end{equation}
In the following we shall assume that the observed body
  is moving according to the integrable dynamics defined by equations
  (\ref{J2eqav}), (\ref{J2eqavshort}), and we shall call oblateness
  effect (or $J_2$ effect) the deviation from the pure Keplerian
  motion defined by these equations.

To solve our problem, we express the equations of motion in terms of
the coordinates $\mathscr{E}^*_{att}$.
First we express the equation \eqref{J2eqavshort} in
Cartesian coordinates $\mathscr{E}_{car} = (\erre,\dot\erre)$.
We obtain
\begin{equation}
\dot{\mathscr{E}}_{car} = {\bf Y}(\mathscr{E}_{car}),
\label{eqcar}
\end{equation}
where 
\begin{align*}
  {\bf Y}
  = \Bigl( \frac{\partial
  \phi_1}{\partial \mathscr{E}_{kep}}{\bf X}_{kep} \Bigr) \circ \phi_1^{-1} 
\end{align*}
is the transformed vector field.
%
From the expression above we obtain the acceleration $\ddot\erre$
as a function of the 
Cartesian coordinates ${\cal E}_{car}$
along the solutions of \eqref{eqcar}:
\begin{equation}
\ddot{\erre} = \Bigl( \frac{\partial \dot\br}{\partial
  \mathscr{E}_{kep}}{\bf X}_{kep} \Bigr) \circ \phi_1^{-1} =:
\tilde{\bf y}(\erre,\erredot) .
\label{J2car}
\end{equation}
\noindent 
As done in \cite{gdbm15} for the pure Keplerian dynamics, we project
the perturbed equation of motion \eqref{J2car} along the line of
sight $\erho$ and obtain the equation
\[
\kepler = 0,
\]
with
\begin{equation}
  \kepler = \Bigl(\ddot\br - \tilde{\bf y} \Bigr)\cdot\erho =
  \ddot{\rho} - \rho\eta^2 + \ddot{\bq}\cdot\erho - \tilde{\bf
    y}\cdot\erho ,
\label{Keq}
\end{equation}
where $\eta = \sqrt{\alphadot^2\cos^2\delta + \deltadot^2}$ is the
proper motion.

Equation \eqref{Keq} can be expressed as a function of the unknown
modified attributable coordinates $(\Delta \alpha,\Delta
\delta,\xi,\zeta)$ using the expressions of $\erre,\dot\erre$ given in
(\ref{geocpos}), (\ref{geocvel}).

\section{The $J_2$ effect on the two-body integrals}
\label{s:intchange}

We recall the expressions of the conserved quantities in the Keplerian
dynamics, i.e. the angular momentum $\angmom$, the energy $\energy$
and the Laplace-Lenz vector $\lenz$, as a function of $\erre,\dot \erre$.
These quantities can be read as functions of the $\elatt$ attributable
coordinates using (\ref{geocpos}), (\ref{geocvel}), and
\begin{eqnarray*}
&&\hskip 0.6cm\vert\dot\br\vert^2 = \xi^2 + \zeta^2 + 2\dqalpha\xi +
  2\dqdelta\zeta + |\rhodot\erho + \dot\bq|^2,\\
&&\hskip 0.6cm\dot\br\cdot\br = \qalpha\xi + \qdelta\zeta +
  (\rhodot\erho + \dot\bq)\cdot\br.
\end{eqnarray*}
We have
\begin{eqnarray*}
&&\hskip 2cm \mathbf{c} = \mathbf A\xi+\mathbf B\zeta+\mathbf C,\\
&&\hskip 2cm\energy = \frac{1}{2}|\erredot|^2 - \frac{\mu}{|\br|},\\
&&\hskip 1cm
\mu\lenz(\rho,\rhodot) = \erredot\times\angmom -
\mu\frac{\br}{|\br|} = \Bigl(\vert\erredot\vert^2
  -\frac{\mu}{|\br|}\Bigr)\br - (\erredot\cdot\br)\erredot,
\end{eqnarray*}
where
\[
\mathbf A = \br\times\ealpha,\quad
\mathbf B = \br\times\edelta,\quad 
\mathbf C = \br\times\dot\bq+
\dot\rho\,\bq\times\erho.
\]

Including the $J_2$ effect in the dynamics the angular momentum and
the Laplace-Lenz vectors are not conserved anymore. However, the
following relations hold true:
\begin{eqnarray}
    R_c\angmom_1 &=& \angmom_2, \label{rotam}\\
    \energy_1 &=& \energy_2, \label{energycons}\\
    R_L\lenz_1 &=& \lenz_2, \label{rotlenz}  
\end{eqnarray}
where
\begin{align*}
R_c        &= R_{\Delta \Omega}^{\bf \hat{z}}, 
\hskip 0.5cm
R_L        = R_{\omega_1 + \Delta \omega}^{\angmom_2} R_{\Delta\Omega}^{\bf \hat{z}} R_{-\omega_1}^{\angmom_1}.
\end{align*}
Here we denote by $R_\phi^{\bv}$ the rotation of an angle $\phi$ around the
axis defined by the vector $\bv$.
The angular variations $\Delta \Omega$ and $\Delta \omega$ are
obtained as
\begin{equation*} \Delta \Omega = \dot \Omega (\tilde{t}_2 - \tilde{t}_1),
    \hskip 1 cm 
    \Delta \omega = \dot \omega (\tilde{t}_2 - \tilde{t}_1),
\end{equation*}
using equation (\ref{J2eqav}) and the epochs $\tilde{t}_i = \bar{t}_i
- \rho_i/c \quad i=1,2$, corrected by aberration, with $c$ the
velocity of light.
\begin{remark}
  We can also write
  \begin{equation*}
    \Delta \Omega = \Omega_2 - \Omega_1, \hskip 1 cm 
    \Delta \omega = \omega_2 - \omega_1,
  \end{equation*}
and
\[
R_L =R_2 R_1^T, 
\]
with
\[
R_j = R_{\omega_j}^{\angmom_j} R_{\Omega_j}^{\bf \hat{z}}, \qquad j=1,2.
\]
\end{remark}

\section{Lambert's theorem with the $J_2$ effect}
\label{s:lambert}

Let us denote by $\lambert$ the expression defining Lambert's
equation.  In the dynamics defined by (\ref{J2eqav}), the mean motion
evolves linearly, thus Lambert's equation can be written as
\begin{equation}
\lambert = n (\tilde{t}_1-\tilde{t}_2)+(\beta - \sin \beta) -
(\gamma-\sin \gamma) + 2k\pi = 0,
\label{lambert}
\end{equation}
with $n$ given by the last equation in (\ref{J2eqav}).
Moreover, $k\in\N$ is the number of revolutions in the time
interval $[\tilde{t}_1,\tilde{t}_2]$.
The angles $\beta$, $\gamma$ are defined by
\begin{equation}
\hskip 0.5cm
\sin^2\frac{\beta}{2} =
\frac{r_1+r_2+d_{L}}{4a}, \qquad
\sin^2\frac{\gamma}{2} =
\frac{r_1+r_2-d_{L}}{4a},
\qquad
0\leq \beta-\gamma\leq 2\pi,
\label{betagamma}
\end{equation}
with $r_1,r_2$ the distances from the center of force.
In (\ref{betagamma}) the distance
\begin{equation}
d_{L} = |\tilde{R}\erre_1 - \erre_2|
\label{rotchord}
\end{equation}
is the length of the chord joining the two positions of the body at
epochs $\tilde{t}_1,\tilde{t}_2$ after rotating the osculating ellipse
at epoch $\tilde{t}_1$ so that it overlaps with the osculating ellipse
at epoch $\tilde{t}_2$.  
The rotation $\tilde{R}$ is given explicitly by
\[
\tilde{R}=\tilde{R}_2 \tilde{R}_1^T, 
\]
with $\tilde{R}_1$ and $\tilde{R}_2$ the transformations from the
selected equatorial reference to the orbital reference frame at the
epochs $\tilde{t}_1$ and $\tilde{t}_2$ respectively (${\bf \hat{n}_1}$
and ${\bf \hat{n}_2}$ are the directions of the lines of nodes):
\[
\tilde{R}_1=R_{\omega_1}^{\bf \hat{c}_1} R_I^{\bf \hat{n}_1} R_{\Omega_1}^{\bf \hat{z}} , 
\hskip 1 cm
\tilde{R}_2=R_{\omega_2}^{\bf \hat{c_2}} R_I^{\bf \hat{n}_2} R_{\Omega_2}^{\bf \hat{z}} .
\]
For a fixed number of revolutions $k$ we have 4 different choices for
the pairs $(\beta, \gamma)$, see \cite{gdbm15} and \cite{battin} for
the details.

\section{Linkage}
\label{s:linkage}

We wish to link two sets of radar data of the form (\ref{radinterp}),
with mean epochs $\bar t_i$, $i=1,2$, and compute one (or more)
preliminary orbits.  In the following we use labels 1, 2 for the
quantities introduced in the previous sections, according to the
epoch.

Moreover, let us define $\bv_2 =\erho_2\times\bq_2$.
Taking into account the $J_2$ effect
we consider the system
\begin{equation}
  (R_c\angmom_1-\angmom_2,\energy_1-\energy_2,\kepler_1,\kepler_2,
  (R_L\lenz_1-\lenz_2)\cdot\bv_2,\lambert) = \bzero
\label{complete}
\end{equation}
of 8 equations in the 8 unknowns $(\bX,\bDelta)$, with
$$
\bX =
(\xi_1, \zeta_1, \xi_2, \zeta_2),
\quad \quad 
\bDelta = (\da_1, \dd_1,
\da_2, \dd_2).
$$
Note that the unknowns are divided into 2 sets so that $\bDelta$ is the
vector of infinitesimal angles.  

We search for solutions of equation
\begin{equation}
{\cal G}(\bDelta) = {\bf G}(\bX(\bDelta),\bDelta) = {\bf 0},
\label{seconde4}
\end{equation}
where
\[
{\bf G} = (\kepler_1,\kepler_2, (R_L\lenz_1-\lenz_2)\cdot\bv_2,{\lambert})
\]
and $\bX(\bDelta)$ is a solution of
\begin{equation}
(R_c\angmom_1-\angmom_2,\energy_1-\energy_2) = {\bf 0}.
\label{prime4}
\end{equation}
In our scheme we use a double-iteration method.
To solve system (\ref{complete}) we use the following procedure.
%
Let us consider the function
\begin{equation}
  {\cal F}(\bDelta) = \frac{1}{2}|{\cal G}(\bDelta)|^2.
  \label{Effe}
\end{equation}
Solutions of (\ref{seconde4}) correspond to absolute minimum
points of (\ref{Effe}). To find them, we search for stationary
points of ${\cal F}$, i.e. solutions of
\begin{equation}
  \frac{\partial{\cal F}}{\partial\bDelta}(\bDelta) =
  \Bigl[\frac{\partial{\cal G}}{\partial\bDelta}(\bDelta)\Bigr]^T{\cal
    G}(\bDelta) = \bzero.
\label{statpts}
\end{equation}
Applying Newton-Raphson method would yield the iterations
\begin{equation}
  \bDelta_{k+1} = \bDelta_k - \Bigl[ \frac{\partial^2{\cal
        F}}{\partial\bDelta^2}(\bDelta_k)\Bigr]^{-1}
  \frac{\partial{\cal F}}{\partial\bDelta}(\bDelta_k), \qquad
  \bDelta_0 = \bzero,
\label{newt_raph_iter}
\end{equation}
where the Hessian matrix
\[
\frac{\partial^2{\cal F}}{\partial\bDelta^2}(\bDelta_k)
\]
has components
\begin{equation}
\frac{\partial^2{\cal F}}{\partial\Delta_j\partial\Delta_i} =
\frac{\partial^2{\cal G}}{\partial\Delta_j\partial\Delta_i}\cdot{\cal G}
+ \frac{\partial{\cal G}}{\partial\Delta_i}\cdot\frac{\partial{\cal G}}{\partial\Delta_j}, \qquad i,j=1,\ldots, 4.
\label{components}
\end{equation}
We drop the terms with the second derivatives of ${\cal G}$, that have
quite complicated expressions. This variation of the algorithm is
usually called {\em pseudo-Newton method}, and can converge if we
start close to a solution. In fact, the second derivatives of ${\cal
  G}$ are multiplied by the components of ${\cal G}$, that are small
in this situation.

At each iteration of the pseudo-Newton method we use again
pseudo-Newton iterations to compute $\bX(\bDelta)$ from system
(\ref{prime4}).  Precisely, for a given value of $\bDelta=\bDelta_k$
we search for $\bX_k = \bX(\bDelta_k)$ by looking for minimum points
of the function
\[
  {\cal H}(\bX) = \frac{1}{2}{\cal J}(\bX)\cdot{\cal J}(\bX),
\]
where
\[
{\cal J}(\bX) =
\left.(R_c\angmom_1-\angmom_2,\energy_1-\energy_2)\right|_{\bDelta=\bDelta_k}.
\]
Taking advantage of the assumed smallness of the solutions $\bDelta$,
we consider $\bDelta=\bzero$ as starting guess.

\begin{remark}
  Equations (\ref{prime4}) are not polynomial in $\bX$, like the
  corresponding equations in \cite{gdbm15}.
\end{remark}

\section{Computing $\bX, \bDelta$}
\label{s:XofDelta}


In equation (\ref{seconde4}) the components of the vector ${\bf
  G}(\bX,\bDelta)$ are similar to the ones of the corresponding
vector in \cite{gdbm15}. However, the following differences occur:
\begin{itemize}
\item[i)] in $\kepler$ at epochs $\tilde{t}_1, \tilde{t}_2$, the term
  $\tilde{\bf y}\cdot\erho$ replaces the radial component of the
  Keplerian force $-\mu\erre\cdot\erho/|\erre|^3$;

\item[ii)] in place of the Laplace-Lenz conservation law we have
  equation (\ref{rotlenz});

\item[iii)] in $\lambert$ we have a different expression of the mean
  motion $n$, as a function of the constant elements $a,e,I$, coming
  from the dynamical model (\ref{J2eqav}). Moreover, we have a
  different expression for the length of the chord, see
  (\ref{rotchord}).
\end{itemize}

To search for the values of $\bDelta$ that solve equation
(\ref{seconde4}) we have to compute the first derivatives of ${\cal
  G}(\bDelta)$ with respect to $\bDelta$, appearing in
(\ref{components}).  Since these computations are similar to the ones
described in \cite[Sect.7]{gdbm15}, we describe below only the
differences coming from the adopted dynamical model.

\subsection{The derivatives of $(R_L\lenz_1-\lenz_2)\cdot\bv_2, \lambert$}

Differentiating $(R_L\lenz_1-\lenz_2)\cdot\bv_2$ with respect to
$\bDelta$ we have to compute the derivatives of the rotation $R_L$.
For this purpose we use Euler-Rodrigues formula for the rotation
matrices (see \cite{gy2015})
\begin{equation}
R_\phi^\bv =  \mathrm{Id} + \sin\phi\hat{\bv} + (1-\cos\phi) \hat{\bv}^2,
\end{equation}
where $\hat{\bv}$ is the skew-symmetric matrix associated to $\bv = (v_1,v_2,v_3)$ by
\[
\hat{\bv}\by = \bv\times\by, \qquad \forall \by\in\R^3,
\]
that is
\[
\R^3\ni {\bf v} \mapsto \hat{\bf v} \stackrel{def}{=}
\left[
\begin{array}{ccc}
0 &-v_3 &v_2\cr
v_3 &0 &-v_1\cr
-v_2 &v_1 &0\cr
\end{array}
\right] .
\]
We have
\begin{equation}
\frac{\partial R_\phi^\bv}{\partial v_j} = \sin\phi\frac{\partial
  \hat{\bv}}{\partial v_j} + (1-\cos\phi)\Bigl(\frac{\partial
  \hat{\bv}}{\partial v_j} \hat{\bv} + \hat{\bv}\frac{\partial
  \hat{\bv}}{\partial v_j} \Bigr), \qquad j=1,2,3.
\end{equation}


\noindent  The derivatives of $\lambert$ differ from the ones written in
 (\cite{gdbm15}) only for the different expressions of $n$ and $d_L$. We have
\begin{eqnarray*}
\frac{\partial n}{\partial \bX} &=& \frac{\partial n_0}{\partial \bX}
\Biggl[1 + \displaystyle \frac{3}{2} J_2
\frac{R^2_{\oplus}}{p^2}\Bigl(1-\frac{3}{2}\sin^2 I\Bigr) \sqrt{1-e^2} \Biggr]\\
&=& -\frac{3}{2\mu}\sqrt{-2\energy_1}
\frac{\partial (2\energy_1)}{\partial \bX}
\Biggl[1 + \displaystyle \frac{3}{2} J_2
  \frac{R^2_{\oplus}}{p^2} \Bigl(1-\frac{3}{2}\sin^2 I\Bigr)
  \sqrt{1-e^2} \Biggr].
\end{eqnarray*} 

\begin{eqnarray*}
\frac{\partial n}{\partial \bE_1} &=& \frac{\partial n_0}{\partial \bE_1}
\Biggl[1 + \displaystyle \frac{3}{2} J_2 \frac{R^2_{\oplus}}{p^2}\Bigl(1-\frac{3}{2}\sin^2 I\Bigr)
  \sqrt{1-e^2} \Biggr]\\
                                  &=& -\frac{3}{2\mu}\sqrt{-2\energy_1}\frac{\partial (2\energy_1)}{\partial  \bE_1}
\Biggl[1 + \displaystyle \frac{3}{2} J_2
\frac{R^2_{\oplus}}{p^2} \Bigl(1-\frac{3}{2}\sin^2 I\Bigr) \sqrt{1-e^2} \Biggr],\\
\frac{\partial n}{\partial \bE_2} &=& {\bf 0}.
\end{eqnarray*}

\subsection{The derivatives of $\kepler_1$, $\kepler_2$}

\noindent We also need to compute the following derivatives:
\[
  \displaystyle \frac{\partial \kepler}{\partial \bDelta} =
  \frac{\partial (\ddot\br\cdot\erho)}{\partial \bDelta} -
  \displaystyle \frac{\partial (\tilde{{\bf y}} \cdot\erho)}{\partial
    \bDelta},
\hskip 1cm
\frac{\partial \kepler}{\partial \bX} = \frac{\partial
  (\ddot\br\cdot\erho)}{\partial \bX} - \displaystyle \frac{\partial
  (\tilde{\bf y} \cdot\erho)}{\partial \bX}.
\]

The derivatives of 
\[
\ddot\br\cdot\erho = \ddot{\rho} - \rho\eta^2 + \ddot{\bq}\cdot\erho
\]
are given by
\begin{align*}
  \frac{\partial (\ddot\br\cdot\erho)}{\partial  \mathscr{E}^*_{att}} &= 
  \begin{bmatrix}
    \ddot{\bq}\cdot\frac{\partial\erho}{\partial\Delta\alpha} & 
    \ddot{\bq}\cdot\frac{\partial\erho}{\partial\Delta\delta} & 
    -\frac{2\xi}{\rho} &
    -\frac{2\zeta}{\rho} &
    -\eta^2 & 0 
  \end{bmatrix}
\end{align*}
with
\begin{equation}
\frac{\partial\erho}{\partial\Delta\alpha} =
\frac{\partial\erho}{\partial\alpha} = \ealpha\cos\delta,
\hskip 1cm
\frac{\partial\erho}{\partial\Delta\delta} =
\frac{\partial\erho}{\partial\delta} = \edelta.
\label{dererho}
\end{equation}
In (\ref{dererho}) we made a little abuse of notation: $\erho$ stands
both for a function of $(\alpha,\delta)$ and for a function of
$(\Delta\alpha,\Delta\delta)$.

\noindent Then, we introduce ${\bf y^*}$ that is the vector $\tilde{\bf
  y}$ as a function of the $\mathscr{E}^*_{att}$ coordinates:
\begin{align*}
  {\bf y^*} = \tilde{\bf y} \circ \phi_2^{-1} \circ \phi_3^{-1} 
  = \Bigl( \frac{\partial \dot\br}{\partial \mathscr{E}_{kep}}{\bf X}_{kep}
\Bigr) \circ \Phi^{-1}.
\end{align*}
The components of ${\bf y^*}$ are
\[
{\bf
  y}_j^* = \Bigl( \frac{\partial \dot\br_j}{\partial \mathscr{E}_{kep}}{\bf
  X}_{kep} \Bigr) \circ \Phi^{-1} \quad j=1,2,3 
\]
and their derivatives are given by
 \begin{align*}
  \frac{\partial {\bf y}_j^*}{\partial \mathscr{E}^*_{att}} &= \Biggl[
    \frac{\partial}{\partial \mathscr{E}_{kep}}\Bigl(\frac{\partial
      \dot\br_j}{\partial \mathscr{E}_{kep}}{\bf X}_{kep}\Bigr) \Biggr]
  \circ \Phi^{-1} \frac{\partial \Phi^{-1}}{\partial
    \mathscr{E}^*_{att}},
\end{align*}
where
\begin{align*}
  \frac{\partial}{\partial \mathscr{E}_{kep}}
\Bigl(\frac{\partial\dot\br_j}{\partial \mathscr{E}_{kep}}
{\bf X}_{kep}\Bigr) &=
  \frac{\partial^2 \dot\br_j}{\partial \mathscr{E}^2_{kep}}{\bf X}_{kep}
   + \frac{\partial \dot\br_j}{\partial \mathscr{E}_{kep}}
    \frac{\partial {\bf X}_{kep} }{\partial \mathscr{E}_{kep}}
\end{align*}
with 
\[
\displaystyle \frac{\partial \dot\br_j}{\partial \mathscr{E}_{kep}} =
\frac{\partial \phi_1^{j+3}}{\partial \mathscr{E}_{kep}}, \quad j=1,2,3\]
and 
\[
\displaystyle \frac{\partial^2 \dot\br_j}{\partial
  \mathscr{E}^2_{kep}} = \frac{\partial^2 \phi_1^{j+3}}{\partial
  \mathscr{E}^2_{kep}}.
\]
Denoting by $(\dot{\Omega}, \dot{\omega}, \dot{\ell})$ the last three
components of the vector field ${\bf X}_{kep}$, see (\ref{J2eqav}), we
have
\begin{equation}
  \frac{\partial {\bf X}_{kep} }{\partial \mathscr{E}_{kep}} =
  \begin{bmatrix}
    {\cal O}_3 &{\cal O}_3 \\
    \frac{\partial(\dot{\Omega}, \dot{\omega}, \dot{\ell})}{\partial(a, e, I)}
    &{\cal O}_3 \\
   \end{bmatrix}
  \label{derXkep}
\end{equation}
where ${\cal O}_3$ is the null matrix of order 3 and the components of
$\frac{\partial(\dot{\Omega}, \dot{\omega}, \dot{\ell})}{\partial(a,
  e, I)}$ are given by
\begin{align*}
  \displaystyle \frac{\partial\dot{\Omega}}{\partial a} &= \displaystyle -\frac{3}{2}J_2 R_{\oplus}^2\cos I
  \frac{\partial [n_0/p^2]}{\partial a}, \qquad\qquad\quad
  \displaystyle \frac{\partial\dot{\Omega}}{\partial e} = \displaystyle -\frac{3}{2}J_2 R_{\oplus}^2 n_0 \cos I 
  \frac{\partial [p^{-2}]}{\partial e}, \qquad\qquad
  \displaystyle \frac{\partial\dot{\Omega}}{\partial I} = \displaystyle  \frac{3}{2}J_2\frac{R_{\oplus}^2}{p^2}n_0\sin I, \\
  \displaystyle \frac{\partial\dot{\omega}}{\partial a} &= \displaystyle  \frac{3}{4}J_2 R_{\oplus}^2(4-5 \sin^2 I) 
  \frac{\partial [n_0/p^2]}{\partial a}, \qquad
  \displaystyle \frac{\partial\dot{\omega}}{\partial e} = \displaystyle  \frac{3}{4}J_2 R_{\oplus}^2 n_0 (4-5 \sin^2 I) 
  \frac{\partial [p^{-2}]}{\partial e}, \qquad
  \displaystyle \frac{\partial\dot{\omega}}{\partial I} = \displaystyle -\frac{15}{2}J_2\frac{R_{\oplus}^2}{p^2}n_0\cos I\sin I,\\
  \displaystyle \frac{\partial\dot{\ell}}{\partial a}   &= \displaystyle  \frac{\partial n_0}{\partial a}+\frac{3}{2}J_2 R_{\oplus}^2 
  \Bigl(1-\frac{3}{2}\sin^2 I\Bigr) \sqrt{1-e^2} \frac{\partial [n_0/p^2]}{\partial a},\\ 
  \displaystyle \frac{\partial\dot{\ell}}{\partial e}   &= \displaystyle  \frac{3}{2}J_2 R_{\oplus}^2 n_0\Bigl(1-\frac{3}{2}\sin^2 I\Bigr)
  \displaystyle \frac{\partial[\sqrt{1-e^2}/p^2]}{\partial e},\\
  \displaystyle \frac{\partial\dot{\ell}}{\partial I}   &= \displaystyle -\frac{9}{2}J_2\frac{R_{\oplus}^2}{p^2}n_0\cos I\sin I\sqrt{1-e^2},
\end{align*}
with
\[
\frac{\partial [n_0/p^2]}{\partial a} = \displaystyle \frac{1}{p^2}\frac{\partial n_0}{\partial a}-\frac{2n_0}{p^3}\frac{\partial p}{\partial a}, \qquad
\frac{\partial [p^{-2}]}{\partial e} = \frac{4ae}{p^3}, \qquad
\frac{\partial[\sqrt{1-e^2}/p^2]}{\partial e} =
\displaystyle \frac{3e}{p^2\sqrt{1-e^2}},
\]
and
\[
\frac{\partial n_0}{\partial a} = -\frac{3}{2}\frac{n_0}{a}, \qquad
\frac{\partial p}{\partial a} = 1-e^2, \qquad
\frac{\partial p}{\partial e} = -2ae.
\]




\section*{Acknowledgments}
This work is partially supported by the Marie Curie Initial Training
Network Stardust, FP7-PEOPLE-2012-ITN, Grant Agreement 317185.



\begin{thebibliography}{}


\bibitem{battin} R.~H. Battin, 1987: An introduction to the
  mathematics and methods of astrodynamics, AIAA Education Series.

\bibitem{b70} R.~A. Broucke, 1970: On the Matrizant of the Two-Body Problem, 
  Astronomy and Astrophysics {\bf 6}, 173--??

\bibitem{CN} A. Celletti, P. Negrini, 1995: Non-integrability of the
  problem of motion around an oblate planet, Celestial Mechanics and
  Dynamical Astronomy, Vol.61, pp.253-260

\bibitem{everhartpitkin1983} E. Everhart and E.T. Pitkin, 1983:
  Universal variables in the two-body problem, American Journal of
  Physics, {\bf 51/8}, 712--717.

\bibitem{ftmr10} D. Farnocchia, G. Tommei, A. Milani, A. Rossi, 2010:
  Innovative methods of correlation and orbit determination for space
  debris, Celest. Mech. Dyn. Astr., Vol. 107(1), pp. 169-185.

\bibitem{gy2015} G. Gallego, A. Yezzi, 2015: A Compact Formula for the
  Derivative of a 3-D Rotation in Exponential Coordinates, Journ. of
  Math. Imaging and Vision {\bf 51/3}, 378--384.

\bibitem{goodyear1965} W.~H. Goodyear, 1965: Completely general
  closed-form solution for coordinates and partial derivative of the
  two-body problem, Astronomical Journal {\bf 70}, 189--??
  
\bibitem{gdm10} G.~F. Gronchi, L. Dimare, A. Milani, 2010: Orbit
  determination with the two-body integrals, Celest. Mech. Dyn. Astr.,
  Vol.107(3), pp. 299-318.

\bibitem{gdbm15} G.~F. Gronchi, L. Dimare, D. Bracali Cioci, H. Ma,
  2015: On the computation of preliminary orbits for Earth satellites
  with radar observations, Monthly Notices of the Royal Astronomical
  Society, Vol.451(2), pp.1883-1891

\bibitem{mg10} A. Milani and G.~F. Gronchi, 2010: Theory of Orbit
  Determination, Cambridge Univ. Press
  
\bibitem{Roy} A.~E. Roy, 2004: Orbital Motion, Fourth Edition, CRC Press.

\bibitem{th1977} L.~G. Taff, D.~L. Hall, 1977: The use of angles and
  angular rates. I: Initial Orbit Determination,
  Celest. Mech. Dyn. Astr. {\bf 16}, 481--488.


\end{thebibliography}
\end{document}